\documentstyle[12pt]{article}
\begin{document}
 
\title{\bf The Limit Cycles of Li\'{e}nard Equations in the
 Strongly Nonlinear Regime}
\author{J.L. L\'{o}pez$^{\dag}$ and R. L\'{o}pez-Ruiz$^{\ddag}$  \\
                                  \\
{$^{\dag}$\small Departamento de Matem\'{a}tica Aplicada} \\
{\small Facultad de Ciencias, Universidad de Zaragoza,
50009-Zaragoza (Spain)} \\
{$^{\ddag}$\small Departamento de F\'{\i}sica Te\'{o}rica} \\
{\small Facultad de Ciencias, Universidad de Zaragoza,
50009-Zaragoza (Spain)}   
\date{ }}

\maketitle
\baselineskip 8mm
 \begin{center} {\bf Abstract} \end{center}
Li\'{e}nard systems of the form $\ddot{x}+\epsilon f(x)\dot{x}+x=0$,
with $f(x)$ an even function, are studied in the strongly nonlinear
regime ($\epsilon\rightarrow\infty$). A method for obtaining the number,
amplitude and loci of the limit cycles of these equations is derived.
The accuracy of this method is checked in several examples.
Lins-Melo-Pugh conjecture for the polynomial case is true in this regime.
$\;$\newline

\newpage
\section{Introduction}

Many systems in nature display self-sustained oscillations:
there is an internal balance between amplification and dissipation,
and they do not require an external periodic forcing to oscillate.
For instance, the beating of a heart, some chemical reactions,
self-excited vibrations in bridges and airplane wings,
B\'{e}nard-von Karman vortex street in the wake of a cylinder, etc.
These phenomena can be modelled by the stable limit cycles found
in specific nonlinear autonomous dynamical systems. These are
called 'self-oscillators' \cite{andronov}.

Limit cycles are isolated closed trajectories in phase space
(an inherently nonlinear phenomenon) \cite{yeyan}.
They describe the periodic
motions of the system. A very well known example having one limit cycle
is the van der Pol equation: $\ddot{x}+\epsilon (x^2-1)\dot{x}+x=0$,
where $\dot{x}(t)=dx(t)/dt$.
It displays a wide range of behavior, from weakly nonlinear to
strongly nonlinear relaxation oscillations when the parameter
$\epsilon$ is modified, making it a good
model for many practical situations \cite{rayleigh,vanderpol,lopez}.
The existence, uniqueness and non-algebraicity
of its limit cycle has been shown 
for the whole range of the parameter $\epsilon$ that controls the
nonlinearity \cite{lienard, lasalle, odani}.

Li\'{e}nard equation,
\begin{equation}
 \ddot{x}+\epsilon f(x)\dot{x}+x=0,
\end{equation}
with $\epsilon$ a real parameter, is a generalization of the van der Pol
self-oscillator. There are no general results about the existence,
number, amplitude and loci of the limit cycles of this system
\cite{hilbert,farkas}.
A nonperturbative method for obtaining information about the
number of limit cycles and their location in phase space when $f(x)$
is an even polynomial is presented in \cite{giacomini}.
 When $f(x)$ is a polynomial of degree
$N=2n+1$ or $2n$, Lins, Melo and Pugh conjectured
(LMP-conjecture) that $n$ is
the maximum number of limit cycles allowed \cite{lins}.
This conjecture is true if $f(x)$ is of degree $2$,
if $f(x)$ is of degree $3$ or if $f(x)$ is even
and of degree $4$. Different results about the
necessary conditions that certain families of $f(x)$ must
satisfy to have $n$ limit cycles
have been given in \cite{lloyd} and references therein.

In this work, we are interested in the strongly nonlinear regime
$(\epsilon\rightarrow\infty)$ of Li\'{e}nard equation when
the viscous term $f(x)$ is a continuos even function,
otherwise arbitrary.
We give a method to find the limit cycles in this regime,
and we claim that LMP-conjecture is true in that limit.

\section{Li\'{e}nard Equation}

We start  by considering the modified form of Li\'{e}nard
equation (1) after the change of variables $\dot{x}(t)=y(x)$ and 
$\ddot{x}(t)=y(x)y'(x)$ (where $y'(x)=dy/dx$):
\begin{equation}
 yy'+\epsilon f(x)y+x = 0.
\end{equation}
The variables are now the coordinates $(x,\dot{x})=(x,y)$
on the plane. Our interest in the shape of limit cycles
in phase space $(x,y)$ justifies the elimination
of the time variable.

\subsection{Symmetries}
A limit cycle $C_l$ of equation (2) is a closed orbit around
the origin $(0,0)$, the only fixed point of the system.
This curve $C_l\equiv (x,y_{\pm}(x))$ cuts the $x$-axis
in two points: $(-a_1,0)$ and
$(a_2,0)$ with $a_1,a_2>0$. Write $(x,y_+(x))$, with
$y_+(x)>0$ and $-a_1<x<a_2$, the positive $y$-branch of the
limit cycle and $(x,y_-(x))$, with $y_-(x)<0$ and $-a_1<x<a_2$,
the negative one. The {\it inversion symmetry}
 $(x,y)\leftrightarrow (-x,-y)$
is verified by equation (2) because $f(x)$ is an even function.
Since the flow lines do not intersect themselves,
a limit cycle and its transformed by this symmetry must be the same curve,
then $y_+(x)=-y_-(-x)$ and $a_1=a_2=a$. Therefore,
in the following, we will restrict ourselves to the positive branches
of limit cycles, $(x,y_+(x))$, with $-a\leq x\leq a$. The amplitude
of oscillation will be the number $a$.
For a given $\epsilon$ this amplitude
$a$ identifies the limit cycle (Fig. 1).
Thus, the number of limit cycles of system (2) is equal
to the number of different possible amplitudes $a$.

Let us remark also the {\it parameter inversion symmetry}
 $(\epsilon,x,y)\leftrightarrow
(-\epsilon,x,-y)$ of equation (2): if $C_l\equiv (x,y_{\pm}(x))$
is a limit cycle for a given $\epsilon$ then $\overline{C}_l\equiv
(x,\overline{y}_{\pm}(x))=(x,-y_{\mp}(x))$ is a limit cycle
for $-\epsilon$. Moreover if
$C_l$ is stable (or unstable) then $\overline{C}_l$ is unstable
(or stable, respectively). This is a consequence
of the fact that each limit cycle encloses the origin $(0,0)$
and this point changes its stability when $\epsilon$
changes of sign. In the regime $\epsilon\rightarrow\infty$
the two eigenvalues of the linear part of system (2) in the origin
are $\lambda_1\sim -\epsilon^{-1} f(0)^{-1}$ and
$\lambda_2\sim -\epsilon f(0)$. In order to have a well defined
stability, we impose the condition $f(0)\neq 0$.
This symmetry is used in our computer-simulations to find
the unstable limit cycles. For a given $\epsilon$ the unstable
limit cycles are the stable ones for $-\epsilon$ after reflection
on the $x$-axis. Summarizing, in order to clasify all the limit
cycles for a given even function $f(x)$, it is enough to find
the positive $y$-branch $y_+(x)$ of the limit cycles of equation (2)
when $\epsilon\rightarrow +\infty$.

\subsection{Scaling}
An easier understanding of the behavior
of equation (2) in the strongly nonlinear regime
is obtained by performing the change of variable $y=\epsilon z$.
With this change Eq. (2) reads:
\begin{equation}
zz'+f(x)z=-\epsilon^{-2}x
\end{equation}

Different scalings can be considered: \newline
On one hand, if we consider $x$ and $f(x)$
of order $1$ in the oscillation region $-a\leq x\leq a$ of
a limit cycle $z(x)$, close to the extreme points (where
$z$ is of order less than $\epsilon^{-2}$),
the limit cycle obeys the equation:
\begin{equation}
zz'+\epsilon^{-2}x = yy'+x=0.
\end{equation}
Integrating this expression we obtain: $x^2+y^2=a^2$, where
$a$ is the amplitude of the limit cycle. If $x=a+\delta x$
and $y=\delta y$, then close to the extreme points $(\pm a,0)$ we have:
$2a\delta x + \delta y^2 = 0$,
with $\delta y\ll\epsilon^{-2}$. \newline
On the other hand, if $z$ is of order bigger than $\epsilon^{-2}$
Eq. (3) is reduced to:
\begin{equation}
z(z'+f(x)) = 0.
\end{equation}
It is clear that all relevant information about the limit cycles
of Eq. (2) when $\epsilon\rightarrow\infty$ is contained
in Eq. (5).

\section{Limit Cycle Solutions}

Local solutions of Eq. (5) are:
\begin{eqnarray*}
z_1(x) & = & 0, \;\;\;\mbox{or} \\
z_2(x) & = & -F(x)+C,
\end{eqnarray*} 
where $F(x)=\int_0^x f(t)dt$ is an odd function and C is a constant.
The positive $y$-branch $y_+$ of a limit cycle
solution of Eq. (1) and 
of amplitude $a$ will be a solution $z_l(x)$ of Eq. (5)
verifying (i) $z_l(-a)=z_l(a)=0$ and (ii) $z_l(0)>0$.
The solution $z_1(x)=0$ for $-a\leq x\leq a$
does not represent a limit cycle because it
does not verify condition (ii). 
The solution $z_2(x)=-F(x)+C$  for $-a\leq x\leq a$
is not a limit cycle either because it 
does not verify condition (i).
Therefore the positive $y$-branch of a limit cycle
(solution of Eq. (5)) must be a piecewise function built
with $z_1(x)=0$ and $z_2(x)=-F(x)+C$ (with $C>0$ by condition (ii))
as integrant blocks.

\subsection{Two-Piecewise Solutions}
We can start trying two-piecewise functions called
{\it two-piecewise limit cycles}.
Write $x_i$ the {\it gluing point} of
the limit cycle $z_l(x)$ where $z_1(x)$ and $z_2(x)$ are {\it glued}:
$z_l(x)$ and $z_l'(x)$ must be continous in $x_i$ because the
velocity $\dot{x}(t)=\epsilon^{-1}z_l(x)$ and the acceleration
$\ddot{x}(t)=\epsilon^{-2}z_l(x)z_l'(x)$ are continuos in the oscillation.
Continuity in $z_l(x)$ means $C=F(x_i)>0$.
Continuity in $z_l'(x)$ means $f(x_i)=0$.
The point $-x_i$ is also a zero of $f(x)$ (in the following
we will suppose $f(x)$ has a finite number of zeroes) and it is
the {\it gluing point} of the negative $y$-branch $y_-$ of the
limit cycle. Thus each pair of zeroes $\pm x_i$ of $f(x)$
can generate at most one limit cycle. If $f(x)$ is a polynomial
of degree $2n$, there will be at most $n$ (two-piecewise)
limit cycles.

The stability of a limit cycle $y(x)$ solution of Eq. (2),
with $-a<x<a$, is determined by the sign of the integral:
\begin{displaymath}
\sigma \equiv -\int_{-a}^{a}\frac{\epsilon f(x)}{y(x)}dx=
-\int_{-a}^{a}\frac{f(x)}{z(x)}dx.
\end{displaymath}
If $\sigma <0$ the limit cycle is stable
and if $\sigma >0$ it is unstable.
In the two-piecewise case this quantity $\sigma$ is controlled
essentially by the region where $z(x)=0$, more exactly  by
the sign of $f(x)$ close to the gluing point $x_i$:
\begin{displaymath}
sign[\sigma] = -sign\left[
\int_{x\sim x_i}\frac{f(x)}{z(x)}dx\right]_{z(x)=0}
= - sign[f(x\sim x_i)]_{z(x)=0}.
\end{displaymath}
In the next section it is shown that if $x_i$ is a gluing point,
$F(x_i)$ must be a maximum of $F(x)$.
Therefore, for $\epsilon>0$, if $x_i\equiv s_i<0$ the limit cycle
whose positive $y$-branch is $y_+(x)=\epsilon z_l(x)$ is stable
because, at the side where $z(x)=0$,
$f(x\sim s_i)>0$. If $x_i\equiv u_i>0$ it is unstable
because, at the side where $z(x)=0$, $f(x\sim u_i)<0$.

Thus the expression 
of the positive $y$-branch $y_+(x)=\epsilon z_i^s(x)$
(solution of Eq. (5)), with $-a_i^s<x<a_i^s$, of
a {\it stable (two-piecewise) limit cycle} is (Fig. 2(a)):
\begin{equation}
z_i^s(x) = \left\{\begin{array}{cl}
0 & \mbox{if}\; -a_i^s<x<s_i \\
-F(x)+F(s_i) & \mbox{if}\;\;\;\;\; s_i<x<a_i^s
\end{array}
\right.
\end{equation}
and the form for an {\it unstable (two-piecewise) limit cycle},
$y_+(x)=\epsilon z_i^u(x)$, with $-a_i^u<x<a_i^u$ is (Fig. 2(b)):
\begin{equation}
z_i^u(x) = \left\{\begin{array}{cl}
-F(x)+F(u_i) & \mbox{if}\; -a_i^u<x<u_i \\
 0 & \mbox{if}\;\;\;\;\; u_i<x<a_i^u,
\end{array}
\right.
\end{equation}
where $a_i^{s,u}>0$ represent the amplitude of each limit cycle,
and $s_i$ and $u_i$ are the gluing points of the
two pieces of each cycle, $z_i^s$ and $z_i^u$, respectively.
Recall that $s_i<0$ for the stable cycle
and $u_i>0$ for the unstable one.
In both cases $F(s_i)>0$, $F(u_i)>0$ and $f(s_i)=f(u_i)=0$.

\subsection{Number of Two-Piecewise Solutions}
\underline{STABLE CYCLES}:
We study in more detail the stable limit cycles given by Eq. (6).
The coordinate
$y_+(x)=\epsilon z_i^s(x)$ of the limit cycle vanishes
in the extreme points 
$\pm a_i^s$ and then $F(s_i)=F(a_i^s)$.
As $z_i^s(x)$
is positive for $s_i<x<a_i^s$ the amplitude $a_i^s$ is defined by:
\begin{displaymath}
a_i^s=\mbox{min}\left\{
x>s_i, F(x)=F(s_i)\right\}.
\end{displaymath}
Also the property $a_i^s>|s_i|$ is fulfilled.

Let us invert the problem. If $s^*<0$ verifies $f(s^*)=0$
and $F(s^*)>0$, is it possible to build a stable
(two-piecewise) limit cycle, as given by Eq. (6), with $s^*$
as gluing point?. First we define $a^*$ by the rule: 
\begin{displaymath}
a^*=\mbox{min}\left\{
 x>s^*, F(x)=F(s^*)\right\}.
\end{displaymath}
Geometrically $a^*$ represents the $x$-coordinate
of the first crossing point between the straight  $z=F(s^*)$ and
the curve $F(x)$ (Fig. 3(a)). If $a^*<|s^*|$
it is not possible to build the limit cycle and we can eliminate this
$s^*$ as a possible gluing point. If $a^*>|s^*|$
the point $s^*$ is a gluing point
candidate. We rename all the pairs  $(s^*,a^*)$
verifying this last property as $(\bar{s}_i,\bar{a}_i^s)$
and collect them into the set:
\begin{equation}
\bar{{\cal A}}^s\equiv\left\{
(\bar{s}_i,\bar{a}_i^s), 
f(\bar{s}_i)=0, F(\bar{s}_i)>0, \bar{s}_{i+1}<\bar{s}_i<0,
\bar{a}_i^s>|\bar{s}_i|
\right\}.
\end{equation}
By construction $\bar{a}_{i+1}^s>\bar{a}_i^s$.

There are two different situations when two sucessive pairs,
$(\bar{s}_i,\bar{a}_i^s)$ and  $(\bar{s}_{i+1},\bar{a}_{i+1}^s)$,
are ordered: \newline
(a) $-\bar{a}_{i+1}^s<\bar{s}_{i+1}<-\bar{a}_i^s<\bar{s}_i$.
In this case it is possible to build a two-piecewise limit cycle
with the pair $(\bar{s}_i,\bar{a}_i^s)$ 
as indicated by Eq. (6). This pair is picked out
and renamed once more as $(s_i,a_i^s)$. \newline
(b) $-\bar{a}_{i+1}^s<-\bar{a}_i^s<\bar{s}_{i+1}<\bar{s}_i$.
Now the constuction of a limit cycle derived from the pair
$(\bar{s}_i,\bar{a}_i^s)$ is not possible. If a initial condition
(with $z=0$) in the interval $[-\bar{a}_i^s,\bar{s}_{i+1}]$ is given,
the system will jump from the point $\bar{s}_{i+1}$ to
$\bar{a}_{i+1}^s>\bar{a}_i^s$, then the curve does not close
at $\bar{a}_i^s$ and there is
no a limit cycle with amplitude $\bar{a}_i^s$. This pair is rejected.

If there is only one pair $(\bar{s}_1,\bar{a}_1^s)$, we consider it
satisfies (a).

All the existing (two-piecewise) stable limit cycles can be found
comparing the pairs $i$ and $i+1$ under rules (a)-(b) and
iterating this process. All the pairs selected by
condition (a) (and renamed as $(s_i,a_i^s)$) are collected into the set:
\begin{equation}
{\cal A}^s\equiv\{(s_i,a_i^s)\}=
\left\{(\bar{s}_i,\bar{a}_i^s)\in\bar{{\cal A}}^s, 
(\bar{s}_i, \bar{a}_i^s)\;\; \mbox{verifies (a)}
\right\}.
\end{equation}
The algorithm above proposed shows also
that $F(x)$ must be a local maximum at each gluing point 
$s_i\in {\cal A}^s$. \newline
The number, $l_s = card({\cal A}^s)$, of pairs
$(s_i,a_i^s)$ is the number of 
stable (two-piecewise) limit cycles of system (5). 

\underline{UNSTABLE CYCLES}:
The same process can be repeated for the unstable cycles
by considering  the points $u^*>0$, where $f(u^*)=0$
and $F(u^*)>0$ and finding their partners
$a^*$ defined as (Fig. 3(b)):
\begin{displaymath}
a^*=\mbox{max}\left\{
x<u^*,
F(x)=F(u^*)\right\}.
\end{displaymath}
Moreover the gluing point candidates must verify
$|a^*|>u^*$. After collecting the pairs
$(u^*,a^*)$ fulfilling this last condition,
we have, as in (8), the set:
\begin{equation}
\bar{{\cal A}}^u\equiv \left\{
(\bar{u}_i,\bar{a}_i^u), 
f(\bar{u}_i)=0, F(\bar{u}_i)>0, \bar{u}_{i+1}>\bar{u}_i>0,
|\bar{a}_i^u|>\bar{u}_i
\right\}.
\end{equation}
A similar algorithm as indicated above can be applied in this case
with the following modified rules: \newline
(a') $\bar{u}_i<-\bar{a}_i^u<\bar{u}_{i+1}<-\bar{a}_{i+1}^u$.
In this case there exists an unstable two-piecewise limit cycle
resulting from the pair $(\bar{u}_i,\bar{a}_i^u)$ and given by Eq. (7).
This pair is picked out and renamed $(u_i,a_i^u)$. \newline
(b') $\bar{u}_i<\bar{u}_{i+1}<-\bar{a}_i^u<-\bar{a}_{i+1}^u$. The
pair $(\bar{u}_i,\bar{a}_i^u)$ does not produce a such limit cycle
and is rejected.

If there is only one pair $(\bar{u}_1,\bar{a}_1^u)$ we consider
it satisfies (a').

We iterate the process given by rules (a')-(b').  
All the pairs selected by condition (a') are collected into
the set:
\begin{equation}
{\cal A}^u\equiv\{(u_i,a_i^u)\}=
\left\{(\bar{u}_i,\bar{a}_i^u)\in\bar{{\cal A}}^u, 
(\bar{u}_i, \bar{a}_i^u)\;\; \mbox{verifies (a')}
\right\}.
\end{equation}
The algorithm above proposed shows also
that $F(x)$ must be a local maximum at each gluing point 
$u_i\in {\cal A}^u$. \newline
The number, $l_u = card({\cal A}^u)$, of pairs $(u_i,a_i^u)$
is the number of unstable (two-piecewise) limit cycles of system (5).
Obviously, $l_s-1\leq l_u\leq l_s+1$.

We claim that the total number $l$ of limit cycles of Eq. (1) in the
strongly nonlinear regime is $l=l_s+l_u$, where $l_s$ and $l_u$
are the number of stable and unstable limit cycles of Eq. (1),
respectively. The amplitudes of these limit cycles are given by
the numbers $a_i^s$ and $a_i^u$, respectively. \newline
We remark also that each pair of zeroes $\pm x_i$
of $f(x)$ produces at most
one limit cycle. If $f(x)$ is a polynomial of degree $2n$ there
will be at most  $n$ limit cycles. Therefore, LMP-conjecture is true
in the strongly nonlinear regime.

\subsection{Shape of the Limit Cycles}

The method introduced above allows us to find the number $l$
of limit cycles and their amplitudes. However,
two-piecewise solutions (Eqs. (6)-(7)) are only
a first approach to the problem of finding the exact loci of
the limit cycles of equation (1) in the
$\epsilon\rightarrow\infty$ regime.
A better approximation to the shape of the limit cycles
is presented in this section.

\underline{STABLE CYCLES}: A stable two-piecewise solution $z_i^s(x)$
of Eq. (5), identified by the pair $(s_i,a_i^s)$, is built of two blocks
glued at $s_i$: $z_{1,i}^s(x)=0$ and
$z_{2,i}^s(x)=-F(x)+F(s_i)$ (see Eq.(6)). The role of
the point $s_i$ is to allow the limit cycle to {\it jump} from the left
side where $z_i^s(x)=z_{1,i}^s(x)=0$
to the maximal amplitude $a_i^s$ through
$z_i^s(x)=z_{2,i}(x)$. Depending of $f(x)$, it is also possible to find
some points inside the interval $-a_i^s<x<s_i$ and
with similar {\it jumping} properties than $s_i$.
If this is the case,
the exact limit cycle in the interval $[-a_i^s,s_i]$
is not $z_i^s(x)=z_{1,i}^s(x)=0$, and the
piece $z_{1,i}^s(x)$ in this 
interval  must be corrected
in the following way (Fig. 4): \newline
(a) Define $\{s_{i,1},\cdots,s_{i,n_i}\}$, the largest set of
points where $F(x)$ is a local maximum and verifying:
\begin{eqnarray}
f(s_{i,1}) = \cdots = f(s_{i,n_i}) = 0, & & \nonumber \\
-a_i^s<s_{i,1}<s_{i,2}<\cdots<s_{i,n_i}<s_i, & & \nonumber \\
0<F(s_{i,1})<F(s_{i,2})<\cdots<F(s_{i,n_i})<F(s_i). & & \nonumber
\end{eqnarray}
The values $\{s_{i,1},\cdots,s_{i,n_i}\}$ will be called
the {\it jumping} points of the cycle $(s_i,a_i^s)$. \newline
(b) For each pair $(i,m)$, $m=1,\cdots,n_i$, we define
$b_{i,m}=\mbox{min}\{x>s_{i,m}, F(x)=F(s_{i,m})\}$.
It follows that $b_{i,m}<s_{i,m+1}$ for every $m$ because
$F(s_{i,m})<F(s_{i,m+1})$. Thus we have the ordered set:
\begin{displaymath}
-a_i<s_{i,1}<b_{i,1}<s_{i,2}<b_{i,2}<\cdots<s_{i,n_i}<b_{i,n_i}<s_i.
\end{displaymath}
(c) The correct expression $z_i^s(x)$ of the stable limit cycle
$(a_i^s,s_i)$ solution of Eq. (5) is:
\begin{equation}
z_i^s(x) = \left\{\begin{array}{cl}
z_{1,i}^s(x) & \mbox{if}\; -a_i^s<x<s_i \\
-F(x)+F(s_i) & \mbox{if}\;\;\;\;\; s_i<x<a_i^s,
\end{array}
\right.
\end{equation}
where
\begin{displaymath}
z_{1,i}^s(x) = \left\{\begin{array}{cl}
0 & \mbox{if}\; -a_i^s<x<s_{i,1} \\
-F(x)+F(s_{i,1}) & \mbox{if}\;\;\;\;\; s_{i,1}<x<b_{i,1} \\
0 & \mbox{if}\;\;\;\;\; b_{i,1}<x<s_{i,2} \\
-F(x)+F(s_{i,2}) & \mbox{if}\;\;\;\;\; s_{i,2}<x<b_{i,2} \\
0 & \mbox{if}\;\;\;\;\; b_{i,2}<x<s_{i,3} \\
\vdots & \vdots\;\;\;\;\;\;\;\;\;\;\;\;\;\;\;\vdots \\
-F(x)+F(s_{i,n_i}) & \mbox{if}\;\;\;\;\; s_{i,n_i}<x<b_{i,n_i} \\
0 & \mbox{if}\;\;\;\;\; b_{i,n_i}<x<s_i.
\end{array}
\right.
\end{displaymath}

Note that the dynamics in the jumping points $s_{i,m}$
has the correct scale to be governed by Eq. (5).
Then the continuity in $z(x)$ and $z'(x)$ can be imposed
in these points in the gluing process. Nevertheless,
a more subtle study precises this gluing process in the points $b_{i,m}$
because other scales not present in Eq. (5) take place in the dynamics.
Thus the continuity of $z'(x)$ (the acceleration)
is lost in the points $b_{i,m}$.
This fact is not a problem in order to consider
the expression (12) as a fine approximation
(up to order $\epsilon^{-2}$) to the shape of
the stable limit cycle, $y_+(x)=\epsilon z_i^s(x)$,
solution of Eq. (1) in the strongly nonlinear regime.

\underline{UNSTABLE CYCLES}: Similarly to the previous case, the piece
$z_{1,i}^u(x)$ in the unstable two-piecewise solution (7) must be corrected.
If the limit cycle is identified by the pair $(u_i,a_i^u)$
the correction in the interval $[u_i,a_i^u]$ is given by the following
steps (Fig. 4): \newline
(a) Define $\{u_{i,1},\cdots,u_{i,p_i}\}$, the largest set of
points where $F(x)$ is a local maximum and verifying:
\begin{eqnarray}
f(u_{i,1}) = \cdots = f(u_{i,p_i}) = 0, & & \nonumber \\
u_i<u_{i,1}<u_{i,2}<\cdots<u_{i,p_i}<a_i^u, & & \nonumber \\
F(u_i)>F(u_{i,1})>F(u_{i,2})>\cdots>F(u_{i,p_i})>0. & & \nonumber
\end{eqnarray}
The values $\{u_{i,1},\cdots,u_{i,p_i}\}$ are
the {\it jumping} points of the cycle $(u_i,a_i^u)$. \newline
(b) For each pair $(i,m)$, $m=1,\cdots,p_i$, we define
$d_{i,m}=\mbox{max}\{x<u_{i,m}, F(x)=F(u_{i,m})\}$.
It follows that $d_{i,m}>u_{i,m-1}$ because
$F(u_{i,m})<F(u_{i,m-1})$. Thus we have the ordered set:
\begin{displaymath}
u_i<d_{i,1}<u_{i,1}<d_{i,2}<u_{i,2}<\cdots<d_{i,p_i}<u_{i,p_i}<a_i^u.
\end{displaymath}
(c) The correct expression $z_i^u(x)$ of the unstable limit cycle
$(a_i^u,u_i)$ solution of Eq. (5) is:
\begin{equation}
z_{1,i}^u(x) = \left\{\begin{array}{cl}
-F(x)+F(u_i) & \mbox{if}\;-a_i^u<x<u_i \\
z_{1,i}^u(x) & \mbox{if}\;\;\;\;\; u_i<x<a_i^u,
\end{array}
\right.
\end{equation}
where
\begin{displaymath}
z_{1,i}^u(x) = \left\{\begin{array}{cl}
0 & \mbox{if}\;\;\;\;\; u_i<x<d_{i,1} \\
-F(x)+F(u_{i,1}) & \mbox{if}\;\;\;\;\; d_{i,1}<x<u_{i,1} \\
0 & \mbox{if}\;\;\;\;\; u_{i,1}<x<d_{i,2} \\
-F(x)+F(u_{i,2}) & \mbox{if}\;\;\;\;\; d_{i,2}<x<u_{i,2} \\
0 & \mbox{if}\;\;\;\;\; u_{i,2}<x<d_{i,3} \\
\vdots & \vdots\;\;\;\;\;\;\;\;\;\;\;\;\;\;\;\vdots \\
-F(x)+F(u_{i,p_i}) & \mbox{if}\;\;\;\;\; d_{i,p_i}<x<u_{i,p_i} \\
0 & \mbox{if}\;\;\;\;\; u_{i,p_i}<x<a_i^u.
\end{array}
\right.
\end{displaymath}

Similar comments to those of the stable case apply here.
Therefore the expression (13) is a fine 
approximation (up to order $\epsilon^{-2}$)
to the shape of the unstable limit cycle, 
$y_+(x)=\epsilon z_i^u(x)$, solution of Eq. (1)
in the strongly nonlinear regime.

\section{Examples}

We illustrate in this section the method introduced in
Sections 2-3 for finding the number, amplitude and loci
of the limit cycles of equation (1) by means of some examples.

{\bf (1)} \underline{$f(x)=5x^4-3x^2-1$}, then $F(x)=x^5-x^3-x$.
The only real solutions of $f(x)=0$ are $x=\pm 0.9157$.
The only local maximum verifying $F(x)>0$ is $s^*=-0.9157$. The
only value $a^*$ verifying $F(a^*)=F(s^*)$ is $a^*=1.3837$.
Moreover, it verifies $a^*>|s^*|$, then we can
rename this pair as $(s^*,a^*)=(\bar{s}_1,\bar{a}_1)$
and because there is only one pair,
$(\bar{s}_1,\bar{a}_1^s)=(s_1,a_1^s)$. The limit cycle when
$\epsilon\rightarrow\infty$ is then $(F(s_1)=1.0397)$:
\begin{displaymath}
z_l(x) = \left\{\begin{array}{cl}
0 & \mbox{if}\;\;\;\;\;\; -1.3857<x<-0.9157 \\
-x^5+x^3+x+1.0397 & \mbox{if}\;\;\;\;\;\; -0.9157<x<1.3857 
\end{array}
\right.
\end{displaymath}
The amplitudes $a_{exp}$ for different values of the parameter
$\epsilon$ can be calculated by numerical integration of Eq. (1).
For instance:
\begin{center}
\begin{tabular}{|c|c|}
\hline
$\epsilon$ & $a_{exp}$ \\
\hline
1 & 1.4099 \\
10 & 1.3975 \\
100 & 1.3874 \\
\hline
\end{tabular}
\end{center}
Let us remark the agreement between our analytical approach and
the behavior of the system.

{\bf (2)} \underline{$f(x)=5x^4-3x^2+0.1$}, then $F(x)=x^5-x^3+0.1x$.
The solutions of $f(x)=0$ are $x_1=\pm 0.7514$ and $x_2=\pm 0.1882$.
The only local maxima verifying
$F(x)>0$ are $s^*=-0.9157$ and $u^*=0.1882$.
In the case of $s^*$ the
only value $a^*$ verifying $F(a^*)=F(s^*)$ is $a^*=1.0045$.
Moreover, it verifies $a^*>|s^*|$, then we can
rename this pair as $(s^*,a^*)=(\bar{s}_1,\bar{a}_1^s)=(s_1,a_1^s)$.
The stable limit cycle, when
$\epsilon\rightarrow\infty$, is then $(F(s_1)=0.1096)$:
\begin{displaymath}
z_l(x) = \left\{\begin{array}{cl}
0 & \mbox{if}\;\;\;\;\;\; -1.0045<x<-0.7514 \\
-x^5+x^3-0.1 x+0.1096 & \mbox{if}\;\;\;\;\;\; -0.7514<x<1.0045 
\end{array}
\right.
\end{displaymath}
The amplitudes $a_{exp}$ for some values of the parameter
$\epsilon$ are:
\begin{center}
\begin{tabular}{|c|c|}
\hline
$\epsilon$ & $a_{exp}$ \\
\hline
1 & 1.0234 \\
10 & 1.0164 \\
100 & 1.0096 \\
\hline
\end{tabular}
\end{center}
In the case of $u^*$ the
closest point $a^*<0$ to the origin verifying $F(a^*)=F(u^*)$
is $a^*=-0.3945$. Moreover, it verifies
 $|a^*|> u^*$, then we can
rename this pair as $(u^*,a^*)=(\bar{u}_1,\bar{a}_1^u)=(s_1,a_1^u)$.
The unstable limit cycle when
$\epsilon\rightarrow\infty$ is then $(F(u_1)=0.0124)$:
\begin{displaymath}
z_l(x) = \left\{\begin{array}{cl}
-x^5+x^3-0.1 x+0.0124 & \mbox{if}\;\;\;\;\; -0.3945<x<0.1882 \\
0 & \mbox{if}\;\;\;\;\;\; 0.1882<x<0.3945 
\end{array}
\right.
\end{displaymath}
The experimental amplitudes $a_{exp}$ for different values
of the parameter $\epsilon$ are:
\begin{center}
\begin{tabular}{|c|c|}
\hline
$\epsilon$ & $a_{exp}$ \\
\hline
1 & 0.3909 \\
10 & 0.3927 \\
100 & 0.3943 \\
\hline
\end{tabular}
\end{center}

{\bf (3)} \underline{$f(x)=-(x^2-0.09)(x^2-0.49)(x^2-0.81)$},
then $F(x)=-0.1428 x^7+0.278 x^5-0.1713 x^3+0.3572 x$.
The solutions of the equation $f(x)=0$ are:
$x=\pm 0.3$, $x=\pm 0.7$ and
$x=\pm 0.9$. The only local maxima are $u_{\alpha}^*=0.9$
and $u_{\beta}^*=0.3$. The value $a_{\alpha}^*=0.5443$ verifies
$F(a_{\alpha}^*)=F(u_{\alpha}^*)$ and
$|a_{\alpha}^*|<u_{\alpha}^*=0.9$
then this pair can not build a limit cycle.
On the other hand, the value $a_{\beta}^*=1.0485$ is the
closest point to the origin verifying $F(a_{\beta}^*)=F(u_{\beta}^*)$.
Moreover, it verifies $|a_{\beta}^*|>u_{\beta}^*$.
Then, we can rename this
pair $(u_{\beta}^*,a_{\beta}^*)=(\bar{u}_1,\bar{a}_1^u)=(u_1,a_1^u)$.
As $u_1^*<|a_1^u|$ the value $u_1^*$ is the jumping
point $u_{1,1}=0.9$ of the cycle $(u_1,a_1^u)$.
The only (unstable) limit cycle of this
system is therefore ($F(u_1)=0.0067$ and $F(u_{1,1})=0.0031$): 
\begin{displaymath}
z_l(x) = \left\{\begin{array}{cl}
0.1428 x^7-0.278 x^5+ & \\
0.1713 x^3-0.3572 x+0.0067 &
\mbox{if}\;\;\;\; -1.0485<x<0.3 \\
0 & \mbox{if}\;\;\;\;\;\; 0.3<x<0.5433 \\
0.1428 x^7-0.278 x^5+ & \\
0.1713 x^3-0.3572 x+0.0031 &
\mbox{if}\;\;\;\;\;\; 0.5433<x<0.9 \\
0 & \mbox{if}\;\;\;\;\;\; 0.9<x<1.0485 
\end{array}
\right.
\end{displaymath}
The experimental amplitudes $a_{exp}$ obtained by direct
integration are:
\begin{center}
\begin{tabular}{|c|c|}
\hline
$\epsilon$ & $a_{exp}$ \\
\hline
500 & 1.0497 \\
1000 & 1.0493 \\
5000 & 1.0489 \\
\hline
\end{tabular}
\end{center}

\section{Conclusions}

Periodic self-oscillations can arise in nonlinear systems.
These are represented by isolated closed curves in phase space
that we call
limit cycles. The knowledge of the number, amplitude
and loci of these solutions in a general nonlinear system is an
unsolved problem.

In this work, we have studied the Li\'{e}nard equation in the strongly
nonlinear regime. An effective algorithm for obtaining
its limit cycles solutions (number, amplitude and loci) has been proposed.
There exists an strong agreement between our analytical approach
and the numerical integration of the system.
Moreover, we claim that Lins-Melo-Pugh conjecture is true in this
regime when the nonlinear viscous term $f(x)$ is an even function.

{\bf Acknowledgements:} We thank Prof. J. Sesma (Zaragoza)
for useful discussions.
We thank also the CICYT (Spanish Governement) for
finantial support.

\newpage
\begin{center} {\bf Figure Captions} \end{center}

{\bf Figure 1}: A typical limit cycle $C_l\equiv (x,y_{\pm}(x))$
of Eq. (1). The oscillatory dynamics $x(t)$ given
by this solution verifies $-a\leq x\leq a$, where
$a$ is the amplitude of oscillation.
Let us observe the inversion symmetry
$(x,y)\leftrightarrow (-x,-y)$ of this curve.

{\bf Figure 2}: Diagrams {\bf (a)-(b)} represent
two-piecewise limit cycle solutions of Eq. (5),
a stable and an unstable, respectively.
Its amplitudes are $a_i^s$ and $a_i^u$. The functions
$z_1(x)$ and $z_2(x)=-F(x)+C$ are glued at the points $\pm s_i$ in the
stable case (a) and at the point $\pm u_i$ in the unstable case (b).

{\bf Figure 3}:  Diagrams {\bf (a)-(b)} show
algorithms for obtaining the 
limit cycles, stable and unstable,
respectively. {\bf (a)} The pair $(s^*,a^*)$
is rejected because $a^*<|s^*|$.
$\bar{{\cal A}}^s =\left\{
(\bar{s}_i,\bar{a}_i^s), i=1,2,3\right\}$.
${\cal A}^s =\left\{(s_3,a_3^s)\right\}$.
{\bf (b)} The pair $(u^*,a^*)$
is rejected because $|a^*|< u^*$.
$\bar{{\cal A}}^u =\left\{
(\bar{u}_i,\bar{a}_i^u), i=1,2,3\right\}$.
${\cal A}^u =\left\{(u_i,a_i^u),i=2,3\right\}$.

{\bf Figure 4}: Shape of
the limit cycles. {\bf (a)} There are four limit cycles:
two stables collected into ${\cal A}^s =\left\{
(s_i,a_i^s), i=1,2,\right\}$, and two unstables into
${\cal A}^u =\left\{
(u_i,a_i^u), i=1,2\right\}$.
{\cal (b)} The unstable cycle $(u_1,a_1^u)$ has the jumping
point $u_{1,1}$. The stable one $(s_2,a_2^s)$ has
the jumping point $s_{2,1}$. Let us remark the {\it repeating}
shape of the four limit cycles.

\end{document}